Response to the Comment by Haack et al. (2015) on the paper by Anfinogenov et al. (2014): John's stone: A possible fragment of the 1908 Tunguska meteorite

Authors:

Yana Anfinogenova,[1]* John Anfinogenov,[2] Larisa Budaeva,[3] and Dmitry Kuznetsov[1]

**Affiliations:**

[1]Yana Anfinogenova, Ph.D., Institute of Physics and Technology, National Research Tomsk Polytechnic University

[2]John Anfinogenov, Tungussky Nature Reserve, Ministry of Natural Resources and Ecology of the Russian Federation

[3]Larisa Budaeva, National Research Tomsk State University

[1]Dmitry Kuznetsov, Ph.D., National Research Tomsk Polytechnic University

*Corresponding author: Dr. Yana Anfinogenova. Address: TPU, 30 Lenin Ave., Tomsk, 634050, Russia. Tel: +79095390220. E-mail: anfiyj@gmail.com and anfy@tpu.ru

**Abstract**

Here we present our response to the Comment on "John's stone: A possible fragment of the 1908 Tunguska meteorite" (Haack et al., 2015, Icarus 265, 238–240). The article provides an open discussion and a critical feedback to the comments of Haack et al. (2015) and emphasizes a significance of the first macroscopic evidence for a candidate meteorite of a new type: planetary-origin meteorite composed of silica-rich sedimentary rock. Discussion concerns the arguments for (i) candidate parental bodies including the Earth, Mars and icy moons of Saturn and Jupiter; (ii) PGE anomaly versus glassy silicate microspherules/quartz grains anomaly in the area of the 1908 Tunguska catastrophe; (iii) isotopic heterogeneity of unmixed silicate reservoirs on Mars; (iv) possible terrestrial loss or contamination in the noble gas signatures in meteorites that spent time in the extreme weather conditions; (v) cosmogenic isotopes and shielding; and (vi) "pseudo meteorites". We conclude that the list of candidate parental bodies for hypothetical sedimentary-origin meteorites includes, but is not limited by the Earth, Mars, Enceladus, Ganymede, and Europa. A parental body should be identified based on the entire body of evidence which is not limited solely by tests of oxygen and noble gas isotopes whose signatures may undergo terrestrial contamination and may exhibit significant heterogeneity within the parental bodies.

**Introduction**

Our team (Anfinogenov *et al.*, 2014; Anfinogenov and Budaeva, 1998) is grateful to authors of the comment (Haack *et al.*, 2015) for their interest in the 1908 Tunguska event and for their work on updating Δ$^{17}$O values and on determining the noble gases isotopic compositions in the samples of John's Stone, an exotic boulder with the numerous signs of high-energy impact in the epicenter of the 1908 Tunguska catastrophe (Anfinogenov *et al.*, 2014).

Briefly, the exotic boulder known as John's Stone has been discovered on the Stoykovich Mountain in the epicenter area of the Tunguska catastrophe (Anfinogenov *et al.*, 2014). Pattern of permafrost destruction suggests high-speed entry and lateral ricochet of John's Stone in the ground with further deceleration and breakage. Landing velocity of John's Stone is estimated to be 547 *m*/s (Anfinogenov *et al.*, 2014). John's Stone is composed of highly silicified gravelite sandstone (~99% $SiO_2$). Outer surface of several splinters of this rock shows continuous glassy coating reminiscent of fresh enamel or fusion crust. There is clear consistency in geometry of Tunguska meteoroid flight trajectory, locations of John's Stone fragments and cleaved pebbles, and directions of impact furrows. John's Stone locates on quaternary deposits at the top of the Stoykovich Mountain and is exotic to the territory within at least hundreds of kilometers around the epicenter (Sapronov, 1986). There are no signs of past glaciation throughout the region (Sapronov, 1986). Decoding of aerial survey photographs covering area within 40 km from the epicenter shows the absence of active diatremes. There is significantly higher (by hundred-fold) content of glassy silicate microspherules precipitated from the atmosphere in the peat layer of 1908 (Dolgov *et al.*, 1973). Macroscopic evidence and data regarding glassy silicate microspherules suggest that John's Stone may be a fragment of the Tunguska meteorite 1908 and may represent a new type of meteorites of planetary origin (Anfinogenov *et al.*, 2014). Carl Sagan said once that "extraordinary claims require extraordinary evidence". We believe that macroscopic signs of impact associated with John's Stone and its nature which is alien to the region of the Tunguska catastrophe represent indeed extraordinary evidence.

However, based solely on the studies of Δ$^{17}$O and noble gases isotopes in the samples of John's Stone, Haack *et al.* (2015) conclude that "John's Stone is a terrestrial rock" and an "unlikely fragment of the projectile responsible for the Tunguska event". We believe it is necessary to raise comments regarding the method and the reasoning Haack *et al.* (2015) used for their attempt for identification of the parental body of John's Stone.

Along with providing an open discussion and a critical feedback to the comments of Haack *et al.* (2015), we would like to emphasize the novelty of our idea and significance of the first macroscopic evidence for a candidate meteorite of a new type. We invite our opponents to reconsider our hypothesis based on the discussion we provide below. Planetary-origin meteorites composed of highly metamorphic sedimentary rocks may indeed exist keeping in mind recent discovery of sedimentary pebble-conglomerate rocks on Mars (Daily Mail Reporter. 2013), the presence of liquid water (Heller *et al.*, 2015; Saur *et al.*, 2015; Steinbrügge *et al.*, 2015; Tanigawa *et al.*, 2014; Vance *et al.*, 2014) and hydrothermal activity within the Solar System (Hsu *et al.*, 2012; Postberg *et al.*, 2011), and possibility of sedimentary rock formation in the presence of liquid water flows generated by tidal forces on the satellites of the giant planets.

**Parental body**

Haack *et al.* (2015) argue that "since sandstones can only form on a parental body with liquid water and, by inference also an atmosphere… there are only two possible parental bodies in the Solar System: the Earth and Mars". Evidence does not support this dictum. Many bodies without a significant atmosphere have abundant liquid water (Heller *et al.*, 2015; Saur *et al.,* 2015; Steinbrügge *et al.,* 2015; Tanigawa *et al.* 2014; Vance *et al.* 2014). Hydrothermal rocks can form on the bodies other than the Earth and Mars (Hsu *et al.*, 2012; Postberg *et al.*, 2011). Moreover, the presence of powerful tidal currents of water in subsurface oceans on icy satellites of Saturn and Jupiter may provide conditions sufficient for formation of sedimentary rocks including sandstones with a variety of grain sizes.

The Saturnian moon Enceladus can be considered a candidate parental body for sedimentary-origin meteorites due to (i) the presence of global hydrothermal activity, (ii) the presence of powerful tidal currents of oceanic water in subsurface oceans potentially resulting in formation of sediments, and (iii) the past history of large-scale impacts explaining ejection of Enceladus' crust fragments into space. Indeed, the plume of Enceladus emits nanometre-sized $SiO_2$ (silica)-containing ice grains (Hsu *et al.,* 2012) formed as frozen droplets from a liquid water reservoir contacting with rock (Postberg *et al.,* 2011). Characteristics of these silica nanoparticles indicate ongoing high temperature (>90 °C) global-scale geothermal and hydrothermal reactions on Enceladus favored by large impacts (Hsu *et al.,* 2015). Enceladus has a differentiated interior consisting of a rocky core, an internal ocean and an icy mantle. Simulation studies suggest that large heterogeneity in the interior, possibly including significant core topography may be due to collisions with large differentiated impactors with radius ranging between 25 and 100 km. Impacts played the crucial role on the evolution of Enceladus (Monteux *et al.,* 2016) and similar effects on evolution are very likely on the other moons of Saturn as well as on other planetary objects, such as Ceres (Davison *et al.,* 2015; Ivanov, 2015).

Recent evidence suggests that a putative subsurface water ocean is present on Ganymede. Iron core of Ganymede is surrounded by a silicate rock mantle and by a globe-encircling, briny subsurface water ocean with alternating layers between high pressure ices and salty liquid water (Saur *et al.,* 2015; Steinbrügge *et al.,* 2015; Vance *et al.,* 2014). If Ganymede or Callisto had acquired their $H_2O$ from newly accreted planetesimals after the Grand Track (Mosqueira and Estrada 2003), then Io and Europa would be water-rich, too (Heller *et al.,* 2015; Tanigawa *et al.,* 2014).

Tidal dissipation and tidal resonance in icy satellites with subsurface oceans are major heat sources for the icy satellites of the giant planets (Kamata *et al.,* 2015). Tidal forces generate heat and currents of liquid water or brine powerful enough to produce sediments that can undergo metamorphic transformations due to hydrothermal activity.

Therefore, there are several candidate parental bodies in the Solar System that may be a place of origin for the sedimentary-type meteorites resembling John's Stone.

**PGE anomaly versus glassy silicate microspherules/quartz grains anomaly**

Haack *et al.* (2015) rule out the Earth as a parental body of John's Stone based on data of Rasmussen *et al.* (1999) and Hou *et al.* (2004) who reported the presence of PGE anomaly in peat cores at an estimated depth of 1908. We believe that the presence of this anomaly could not rule out the possibility of an independent event or the presence of another anomaly at the same area, essentially, based on the laws of logic. As a matter of fact, PGE anomaly is not the only anomaly known in the region of the 1908 Tunguska catastrophe.

In our original article (Anfinogenov *et al.,* 2014) we cited paper reporting the discovery of significantly higher content of glassy silicate microspherules precipitated from the atmosphere in peat layer of 1908 throughout the area of the Tunguska catastrophe (Dolgov *et al.* 1973). Peat layer of 1908 contained up to hundredfold-higher count of gray and colorless transparent silicate microspherules than the adjacent peat layers. Data of neutron activation analysis showed that chemical composition of microspherules was distinct from that of industrial glass, known stony meteorites, tektites, and Moon rocks (Kolesnikov *et al.,* 1976).

It is essential to note that Dr. E.M. Kolesnikov who performed the aforementioned neutron activation analysis (Kolesnikov *et al.,* 1976) of anomalous silicate microspherules from peat layer of 1908 is also a co-author of both papers by Rasmussen *et al.* (1999) and by Hou *et al.* (2004) reporting PGE anomaly. Therefore, two anomalies are present in the area of interest and they both are reported by the same Tunguska meteorite explorer, Dr. E.M. Kolesnikov: (i) PGE anomaly and (ii) anomalous abundance of glassy silicate microspherules precipitated from the atmosphere in peat layer of 1908. Even more so, the quartz grains were found in sediment cores collected from Lake Cheko (Gasperini *et al.*, 2009) suggesting that they might have resulted from dust produced by the explosion in the atmosphere of the main body if the Tunguska cosmic body were silica-rich.

No macroscopic pieces of chondritic or cometary projectile have been indeed found in the area of the 1908 Tunguska catastrophe. In 1993, Chyba *et al.* reported in Nature magazine that carbonaceous asteroids and especially comets are unlikely candidates for the Tunguska object. The Tunguska event represents a typical fate for stony asteroids tens of meters in radius entering the Earth's atmosphere at common hypersonic velocities (Chyba *et al.,* 1993). In this regard, John's Stone represents a sound macroscopic candidate for a stony impactor though of a previously unknown type. This hypothesis is consistent with John's Stone phenomenon bearing the numerous signs of high-energy impact and glassy fusion crust-like surface modification on some splinters. It is also consistent with the discovery of the quartz grains in sediment cores collected from Lake Cheko (Gasperini *et al.,* 2009) and the glassy silicate microspherules anomaly associated with the area of the 1908 Tunguska catastrophe (Dolgov *et al.,* 1973; Kolesnikov *et al.,* 1976).

Studies by Rasmussen *et al.* (1999) and by Hou *et al.* (2004) invoke questions regarding the methodology. These studies are based on data from an insignificant number of peat core samples showing PGE anomaly: only four adjacent peat columns and only one peat column were studied by Rasmussen *et al.* (1999) and by Hou *et al.* (2004), respectively. Peat cores from singular topographic locations were tested in these studies. On the contrary, Dolgov *et al.* (1973) described anomalous content of glassy silicate microspherules in peat layer of 1908 in hundreds of peat samples associated with the entire area of the 1908 Tunguska catastrophe.

Two explanations may be proposed for the presence of these two anomalies (PGE and glassy silicate microspherules/quartz grains) in the area of the 1908 Tunguska catastrophe: (i) occurrence of two impact events independently involving silica-rich impactor and chondritic or cometary projectile or (ii) complex conglomerate composition of the impactor consisting of multiple parts that merged due to either collision of parental asteroids in outer space or due to a co-ejection of different adjacent rocks from a parental planetary body due to a high-energy impact. For example, it could be a co-ejection of enclosing bedrocks, intrusive igneous rocks, and impactor material where any of these components could partially melt due to impact. These processes could produce so-called rubble-pile asteroids.

Therefore, neither the Earth nor other bodies in the Solar System can be ruled out as candidate parental bodies solely on the ground of PGE anomaly reported by Rasmussen *et al.* (1999) and Hou *et al*. (2004). Moreover, the anomaly of glassy silicate microspherules/quartz grains (Dolgov *et al.,* 1973; Gasperini *et al*., 2009) provides a statistically significant evidence for silica-rich projectile responsible for the 1908 Tunguska event.

**Isotope tests**

Haack *et al.* (2015) rule out Mars as a candidate parental body for John's Stone based on oxygen and noble gas isotope tests. It is necessary to emphasize that isotopic characterizations *per se* are very essential and further elemental and isotopic characterizations of this exotic Tunguska boulder are highly encouraged. However, numerical results of isotopic characterizations though important on their own do not prove the conclusions of Haack *et al.* (2015).

No rock samples have ever been delivered from Mars to the Earth before. Isotopic compositions of Martian rocks have never been tested by Mars rovers yet. All which was tested up to day were the so called Martian meteorites. However, diversities in the rock-forming processes and in the corresponding rock types within planets are great and insufficiently studied. A significant heterogeneity in $\Delta^{17}O$ in rocks of different types has been reported (Wang *et al.*, 2013). Pack and Herwartz (2015; 2014) provide evidence that the concept of a single terrestrial mass fractionation is invalid on small-scale. They conclude that mineral assemblages in rocks fall on individual "rock" mass fractionation lines with individual slopes and intercepts.

We believe the same may be true for other bodies of the Solar System such as Mars, large moons of Jupiter and Saturn, and large asteroids. An idea of separate long-lived silicate reservoirs on Mars is supported by radiogenic isotope studies (Borg *et al.*, 1997; 2003). The distinct $\Delta^{17}O$ and $\delta^{18}O$ values of the silicate fraction of NWA 7034 compared to other SNC meteorites supports the idea of distinct lithospheric reservoirs on Mars that have remained unmixed throughout Martian history (Agee *et al.*, 2013; Ziegler *et al.*, 2013). Isotopic heterogeneity, including that in the noble gases, can be significant in the Martian mantle. Models for accretion and early differentiation of Mars were tested with chronometers several of which provided evidence of very early isotopic heterogeneity preserved within Mars (Halliday *et al.*, 2001). If significant heterogeneities are reported for known Martian meteorites which are all igneous rocks, then even greater isotopic heterogeneities may exist for the rocks of different types within planet.

The most essential issue is that the work of Haack *et al.* (Haack *et al.*, 2015) did not provide any characterizations of any particular terrestrial and extraterrestrial rocks **similar** to John's Stone (highly metamorphic highly silicified gravelite sandstones with $SiO_2$ content of nearly 99%), which would be helpful for methodologically proper justification of the conclusions they drew. The tests, performed with the samples of John's Stone, should be done with **similar** extraterrestrial rocks which may be feasible midterm considering that matching sedimentary silica-rich rocks have been found on Mars (Bandfield *et al.*, 2004; 2006; Christensen *et al.*, 2005; Edgett and Malin 2000; Jerolmack, 2013; Kerber and Head, 2012; McLennan, 2003; Michalski, 2013; Smith and Bandfield, 2012; Squyres *et al.*, 2008; Williams *et al.*, 2013). Indeed, recent studies showed that there are sedimentary rocks on Mars including pebble conglomerates and sandstones (Williams *et al.*, 2013; Kerber and Head, 2012). Quartz-bearing deposits are consistently co-located with hydrated silica on Mars (Smith and Bandfield, 2012). There is striking similarity between Martian pebble conglomerates (images are available at Daily Mail Reporter 2013) and John's Stone (Anfinogenov *et al.*, 2014). Notably, the Tunguska boulder, in addition to $SiO_2$, contains traces of a Ti-oxide phase and its bulk composition (Bonatti *et al.*, 2015) is similar to the composition inferred from APXS data (Squyres *et al.*, 2008) for the silica deposits of the Gusev crater on Mars. A paper by Bonatti *et al.* (2015) contains a section with detailed review of sedimentary rocks and hydrothermal activity on Mars.

On the other hand, terrestrial rocks reminiscent of John's Stone would represent necessary terrestrial control required to be tested in the same set of experiments by using the same equipment. All the more so because the difference between estimations of $\Delta^{17}O$ by Bonatti *et al.* (2015) and by Haack *et al.* (2015) may suggest the presence of equipment-specific errors or human factors.

Instead of comparing John's Stone with **similar** terrestrial and extraterrestrial rocks, Haack *et al.* (2015) compared isotopic compositions of John's Stone with published data on Chelyabinsk meteorite (Nishiizumi *et al.*, 2013; Buzemann *et al.*, 2014), SNC meteorites (Franchi *et al.*, 1999), NWA 7034 (Agee *et al.*, 2013), CIs (Clayton and Mayeda, 1999), Lunar rocks (no reference was provided by Haack *et al.*, 2015), angrites and HEDs (Greenwood *et al.*, 2005), aubrites and enstatite chondrites (Newton *et al.*, 2000), and averaged characteristics of unidentified quartz-rich terrestrial rocks. This comparison is valuable on its own, but, due to unique occurrence of the Tunguska boulder, it does not prove the conclusions of Haack *et al.* (2015).

Therefore, we strongly suggest more careful consideration of Mars as a candidate parental body of John's Stone.

**Possible terrestrial loss or contamination in the noble gas signatures**

Terrestrial loss or contamination in the noble gas signatures should be considered for any meteorite including Martian meteorites that spent time in terrestrial environment (Schwenzer *et al.*, 2013). All the more so for Tunguska meteorite because weather conditions in Tunguska are extreme and temperatures range from –61 °C during winter to +40 °C during summer. If John's Stone is a fragment of the Tunguska impactor, then the repeated freezing of water easily penetrating into tiny pores in its material for decades could significantly accelerate contamination of the rock with terrestrial elements concealing original isotopic signatures.

**Shielding**

Haack *et al.* (2015) state that "The lack of any cosmogenic noble gases (particularly striking in $^3$He, $^{21}$Ne, $^{38}$Ar) would be consistent with an extraterrestrial origin under large shielding". We agree with this notion. Indeed, a pre-atmospheric diameter of Tunguska cosmic body is estimated to be tens of meters in diameter (Chyba *et al.*, 1993) suggesting up to 50 m of shielding whereas a core part of the Tunguska impactor would have better chances of surviving passage through the atmosphere.

**Pseudo meteorites**

We found a total of 226 records for meteorites with status of "pseudo" in the category of "terrestrial meteorites" in the Meteoritical Bulletin Database of the International Society for Meteorites and Planetary Science (The Meteoritical Society). Except NWA 6944, all other "pseudo" meteorites were found in Antarctica. No characteristics are provided for description of these rocks, but the fact that these specimens were considered candidates for meteorites and were included in the Meteoritical Bulletin Database might suggest that their minerology was exotic to the areas of discovery and/or they presented with a glassy cover or a fusion crust-like surface. It would be helpful to know more about those rocks. Sedimentary-origin meteorites may be present among them.

**Technical issues with representation of the noble gas signatures reported by Haack *et al.* (2015)**

We also would like to point out some inconsistencies in data published by Haack *et al.* (2015). In particular, Table 1 shows that $^{36}$Ar was not detected in four out of five samples of John's Stone. This absence of $^{36}$Ar would agree with data generated by Mars rover suggesting a depletion of Martian atmosphere in light Ar isotopes (Atreya *et al.,* 2013). However, for some reason, Haack *et al.* (2015) state that the ratio of light to heavy argon isotopes in the samples of John's Stone corresponds to the terrestrial one. Table 1 (Haack *et al.,* 2015) does not suggest that. We do not understand how five ratios of $^{36}$Ar/$^{40}$Ar were calculated when $^{36}$Ar was not detected in four out of five samples. $^{38}$Ar contents cannot be understood from Table 1 for the same reason.

**Summary**

The numerical results of the isotopic tests performed by Haack *et al.* (2015) present valuable phenomenological information regarding the Tunguska boulder historically called John's Stone. However, the conclusions of Haack *et al.* (2015) regarding the identification of its parental body do not directly follow from the results of their tests. These conclusions represent an opinion rather than an interpretation which is, in our view, insufficiently evidence-based and sounds like "extraterrestrial rocks of this particular isotopic composition do not exist because they could not exist" according to the scientific dictum. Moreover, while drawing the conclusions, Haack *et al.* (2015) ignored **the entire pool of evidence** suggesting high-energy impact associated with the John's Stone phenomenon described by Anfinogenov *et al.* (2014).

Authors of the original article (Anfinogenov *et al*., 2014) invite scientific community to consider the significance of our hypothesis for the extraterrestrial origin of John's Stone found with signs of high-energy impact in the epicenter of the 1908 Tunguska catastrophe. Further in-depth study of this rock should be undertaken including the thermoluminescence analysis, rock age determination, and the comparison of John's Stone signatures with those of **similar** terrestrial and extraterrestrial rocks. The site of impact associated with John's Stone requires comprehensive interdisciplinary field examination.

We conclude that the list of candidate parental bodies for hypothetical sedimentary-origin meteorites includes, but is not limited by the Earth, Mars, Enceladus, Ganymede, and Europa. A parental body should be identified based on consensus of all evidence, not limited solely by tests for oxygen and noble gas isotopes which may undergo terrestrial contamination and exhibit significant heterogeneity within the parental bodies. The study of the Solar System is just in the beginning.

**Acknowledgments**

We thank Dr. Haack and two anonymous reviewers who provided critical feedback in a framework of private discussion in Icarus that allowed us to clarify our viewpoints and improve linguistics of the manuscript.